\documentclass[12pt,english]{article}
\pdfoutput=1
\usepackage{graphicx,array}
\usepackage{cite}
\usepackage{color}
\usepackage{latexsym}
\usepackage{amsthm}
\usepackage{amsmath}
\usepackage[titletoc]{appendix}
\usepackage{enumitem}
\usepackage{amssymb}
\usepackage{hyperref}
\usepackage[hang,flushmargin]{footmisc} 
\usepackage{stmaryrd}
\usepackage{relsize}

\usepackage{epsfig}
\usepackage{caption}
\usepackage{subcaption}
\usepackage{float}

\usepackage{slashed}
\usepackage{mwe}
\usepackage{pifont}
\usepackage{feynmp}
\usepackage{multirow}
\def\simgt{\mathrel{\lower2.5pt\vbox{\lineskip=0pt\baselineskip=0pt
           \hbox{$>$}\hbox{$\sim$}}}}
\def\simlt{\mathrel{\lower2.5pt\vbox{\lineskip=0pt\baselineskip=0pt
           \hbox{$<$}\hbox{$\sim$}}}}

\newcommand{\be}{\begin{equation}}
\newcommand{\ee}{\end{equation}}
\newcommand{\bea}{\begin{eqnarray}}
\newcommand{\eea}{\end{eqnarray}}
\newcommand{\Eq}[1]{Eq.~(\ref{#1})}
\newcommand{\Eqs}[2]{Eqs.~(\ref{#1}) and (\ref{#2})}
\newcommand{\Sec}[1]{Sec.~\ref{#1}}

\newcommand{\Fig}[1]{Fig.~\ref{#1}}

\newcommand{\Ref}[1]{Ref.~\cite{#1}}

\newcommand{\ab}[1]{\langle #1 \rangle}

\newcommand{\lrbraces}[1]{\lbrace #1 \rbrace}

\newcommand{\Ms}{\widetilde{\cal M}}

\newcolumntype{M}[1]{>{\centering\arraybackslash}m{#1}}

\newcommand*\oline[1]{%
  \vbox{%
    \hrule height 0.5pt
    \kern0.68ex
    \hbox{%
      \kern-0.1em
      \ifmmode#1\else\ensuremath{#1}\fi
      \kern-0.1em
    }
  }
}

\definecolor{nicered}{rgb}{0.7,0.1,0.1}
\definecolor{nicegreen}{rgb}{0.1,0.5,0.1}
\hypersetup{
 colorlinks,citecolor=black,linkcolor=black,urlcolor=black}

\begin{document}
\interfootnotelinepenalty=10000
\baselineskip=18pt
\hfill CALT-TH-2015-005
\hfill

\vspace{2cm}
\thispagestyle{empty}
\begin{center}
{\LARGE\bf
Simple Recursion Relations \\ \medskip for General Field Theories
}\\
\bigskip\vspace{1.cm}{
{\large Clifford Cheung\textasciicircum , Chia-Hsien Shen\textasciicircum, and Jaroslav Trnka\textasciicircum}
} \\[7mm]
 {\it \textasciicircum Walter Burke Institute for Theoretical Physics, \\[-1mm]
    California Institute of Technology, Pasadena, CA 91125}\let\thefootnote\relax\footnote{e-mail: \url{clifford.cheung@caltech.edu}, \url{chshen@caltech.edu}, \url{trnka@caltech.edu}} \\
 \end{center}
\bigskip
\centerline{\large\bf Abstract}

\begin{quote} \small
On-shell methods offer an alternative definition of quantum field theory at tree-level, replacing Feynman diagrams with recursion relations and interaction vertices with a handful of seed scattering amplitudes.  In this paper we determine the simplest recursion relations needed to construct a general four-dimensional quantum field theory of massless particles.   For this purpose we define a covering space of recursion relations which naturally generalizes all existing constructions, including those of BCFW and Risager.  
The validity of each recursion relation hinges on the large momentum behavior of an $n$-point scattering amplitude under an $m$-line momentum shift, which we determine solely from dimensional analysis, Lorentz invariance, and locality. We show that all amplitudes in a renormalizable theory are 5-line constructible. Amplitudes are 3-line constructible if an external particle carries spin or if the scalars in the theory carry equal charge under a global or gauge symmetry.  Remarkably, this implies the 3-line constructibility of all gauge theories with fermions and complex scalars in arbitrary representations, all supersymmetric theories, and the standard model.  Moreover, all amplitudes in non-renormalizable theories without derivative interactions are constructible; with derivative interactions, a subset of amplitudes is constructible. We illustrate our results with examples from both renormalizable and non-renormalizable theories.   Our study demonstrates both the power and limitations of recursion relations as a self-contained formulation of quantum field theory.
\end{quote}

\setcounter{footnote}{0}

\newpage
\tableofcontents

\newpage


\section{Introduction}

On-shell recursion relations are a powerful tool for calculating
tree-level scattering amplitudes in quantum field theory.  Practically, they are far more efficient than Feynman diagrams.  Formally, they offer hints of an alternative boundary formulation of quantum field theory grounded solely in on-shell quantities.  To date, there has been enormous progress in computing tree-level scattering amplitudes in various gauge and gravity theories with and without supersymmetry.  

In this paper we ask: {\it to what extent do on-shell recursion relations define quantum field theory?}  Conversely,  for a given quantum field theory, what is the minimal recursion relation, if any, that constructs all of its amplitudes?  Here an amplitude is ``constructible'' if it can be recursed down to lower point amplitudes, while a theory is ``constructible" if all of its amplitudes are either constructible or one of a finite set of seed amplitudes which initialize the recursion.  

For our analysis we define a ``covering space'' of recursion relations, shown in \Eq{eq:genshift}, which includes natural generalizations of the BCFW~\cite{Britto:2005fq} and Risager~\cite{Risager:2005vk} recursion relations.  These generalizations, defined in \Eq{eq:mline} and \Eq{eq:m1line}, intersect at a new ``soft'' recursion relation, defined in \Eq{eq:softshift}, that probes the infrared structure of the amplitude.

As usual, these recursion relations rely on a complex deformation of the external momenta parameterized by a complex number $z$.  By applying Cauchy's theorem to the complexified amplitude, ${\cal M}(z)$, one relates the original amplitude to the residues of poles at complex factorization channels, plus a boundary term at $z=\infty$ which is in general incalculable.  Consequently, an amplitude can be recursed down to lower point amplitudes if it vanishes at large $z$ and no boundary term exists. 

The central aim of this paper is to determine the conditions for on-shell constructibility by determining when the boundary term vanishes for a given amplitude.   We define the large $z$ behavior, $\gamma$, of an amplitude by
\bea {\cal M}(z\rightarrow \infty) = z^{\gamma},
\label{eq:gamma}
\eea
for an $n$-point amplitude under a general $m$-line momentum shift, where $m\leq n$.  Inspired by~\Ref{ArkaniHamed:2008yf}, we rely crucially on the fact that the large $z$ limit describes the scattering of $m$ hard particles against $n-m$ soft particles.  Hence, the large $z$ behavior of the $n$-point amplitude is equal to the large $z$ behavior of an $m$-point amplitude computed in the presence of a soft background.  Fortunately,  explicit $m$-point amplitudes need not be computed, as $\gamma$ can be stringently bounded simply from dimensional analysis, Lorentz invariance, and locality, yielding the simple formulas in \Eq{eq:bestbound}, \Eq{eq:bound1}, \Eq{eq:bound_h}, and \Eq{eq:bound_mixed}.   From these large $z$ bounds, it is then possible to determine the minimal $m$-line recursion relation needed to construct an $n$-point amplitude for any given theory.   If every amplitude, modulo the seeds, are constructible, then we define the theory to be $m$-line constructible.

\begin{table}[t]
	\centering
	\begin{tabular}{c ||c | c | c | c | c | c | c | c|c|}
		 Theory & YM & YM + $\psi$ & YM + $\phi$ &   YM + $\psi$ + $\phi$  & Yukawa & Scalar & SUSY & SM  \\
		\hline \hline
		$m$ & 2 & 2 & 5 (3) & 5 (3) & 3& 5 (3) & 3 & 3
	\end{tabular}
	\caption{Summary of the minimal $m$-line recursion relation needed to construct all scattering amplitudes in various renormalizable theories: Yang-Mills with matter of diverse spins and arbitrary representations, Yukawa theory, scalar theory, supersymmetric theories, and the standard model.  The values in parentheses apply if every scalar has equal charge under a $U(1)$ symmetry. Here $\phi$ and $\psi$ denote scalars and fermions, respectively.}
	\label{table:renormalizable_summary}
\end{table}

Our results apply to a general quantum field theory of massless particles in four dimensions, which we now summarize as follows:
\medskip

\noindent {\bf Renormalizable Theories}
\begin{itemize}
\item Amplitudes with arbitrary external states are $5$-line constructible.
\item Amplitudes with any external vectors or fermions are $3$-line constructible.  
\item Amplitudes with only external scalars are $3$-line constructible if there is a $U(1)$ symmetry under which every scalar has equal charge.  

\item The above claims imply $5$-line constructibility of all renormalizable quantum field theories and $3$-line constructibility of all gauge theories with fermions or complex scalars in arbitrary representations, all supersymmetric theories, and last but not least the standard model.  The associated recursion relations are defined in \Eq{eq:mline} and \Eq{eq:m1line}.
\end{itemize}

\noindent {\bf Non-renormalizable Theories}
\begin{itemize}
\item  Amplitudes are $m$-line constructible for $(m-1)$-valent interactions without derivatives.
\item  Amplitudes are constructible for interactions with derivatives up to a certain order in the derivative expansion.  

\item The above claims imply $m$-line constructibility of all scalar and fermion $\phi^{m_1}\psi^{m_2}$ theories for $m_1+m_2=m-1$, and of certain amplitudes in higher derivative gauge and gravity theories.  The associated recursion relations are defined in \Eq{eq:genshift}.
\end{itemize}
Constructibility conditions for some familiar cases are presented in Tab.~\ref{table:renormalizable_summary}. These cases fully span the space of all renormalizable theories.

As we will see, our covering space of recursion relations naturally bifurcates according to the number of $z$ poles in each factorization channel: one or two.  For the former, the recursion relations take the form of standard shifts such as BCFW and Risager, which is the case for the 5-line and 3-line shifts employed for renormalizable theories.  For the latter, the recursion relations take a more complicated form which is more cumbersome in practice, but necessary for some of the non-renormalizable theories.

 The remainder of our paper is as follows.  In \Sec{sec:cover}, we present a covering space of recursion relations for an $m$-line shift of an $n$-point amplitude, taking note of the generalizations of the BCFW and Risager momentum shifts.  Next, we compute the large $z$ behavior for these momentum shifts in \Sec{sec:z}.   Afterwards, in \Sec{sec:th} we present our main result, which is a classification of the minimal recursion relations needed to construct various renormalizable and non-renormalizable theories.  Finally, we discuss examples in \Sec{sec:ex} and conclude in \Sec{sec:outlook}.
 
\section{Covering Space of Recursion Relations}
\label{sec:cover}

\subsection{Definition}

Let us now define a broad covering space of recursion relations subject to a loose set of criteria.
In particular, we demand that the external momenta remain on-shell and conserve momenta for all values of $z$.  In four dimensions, these conditions are automatically satisfied if the momentum deformation is a complex shift of the holomorphic and anti-holomorphic spinors of external legs\footnote{There is a more general class of shifts in which both $\lambda_i$ and $\widetilde\lambda_i$ are shifted for every particle.  However, in the case momentum conservation imposes complicated non-linear relations among reference spinors which makes the study of large $z$ behavior difficult.},
\bea
 \lambda_i &\rightarrow& \lambda_i(z)= \lambda_i + z\eta_i , \quad i \in {\cal I} \nonumber\\
\widetilde \lambda_i &\rightarrow& \widetilde\lambda_i(z)= \widetilde \lambda_i + z\widetilde \eta_i, \quad i \in \widetilde{\cal I}, 
\label{eq:genshift}
\eea
where $\eta_i$ and $\widetilde \eta_i$ are reference spinors that may or may not be identified with those of external legs, and ${\cal I}$ and $\widetilde{\cal I}$ are disjoint subsets of the external legs. As shorthand, we will refer to the shift in \Eq{eq:genshift} as an $[\widetilde{\cal I},{\cal I}\rangle$-line shift.  When the specific elements of ${\cal I}$ and $\widetilde{\cal I}$ are not very important, we will sometimes refer to this as an $[|\widetilde {\cal I}| , |{\cal I}|\rangle$-line shift, where the labels are the orders of ${\cal I}$ and $\widetilde{\cal I}$.  For an $m$-line shift, $m=|{\cal I}| + |\widetilde{\cal I}|$.  In this notation, the BCFW and Risager shifts are $[1,1\rangle$-line and $[3,0\rangle$-line shifts, respectively.

As we will see, the efficacy of recursion relations depend sensitively on the correlation between the helicity of a particle and whether its holomorphic or anti-holomorphic spinor is shifted.  Throughout, we will define ``good'' and ``bad'' shifts according to the choices
\bea
({\cal I}, \widetilde{\cal I}) &=& \left\{ \begin{array}{ll}
(+,-),&\quad  \textrm{good shift}\\
(-,+),&\quad  \textrm{bad shift}
\end{array}
\right.
\label{eq:goodbad}
\eea
For example, the bad shift for the case of BCFW yields a non-vanishing contribution at large $z$ in non-supersymmetric gauge theories.

The resulting tree amplitude, ${\cal M}(z)$, is then complexified, but the original amplitude, ${\cal M}(0)$ is obtained by evaluating the contour integral $\oint dz\, {\cal M}(z)/z $ for a contour encircling $z=0$.   An on-shell recursion relation is then obtained by applying Cauchy's theorem to deform the contour out to $z=\infty$, in the process picking up all the residues of ${\cal M}(z)$ in the
complex plane.

As noted earlier, the momentum conservation must apply for arbitrary values of $z$, implying
\bea \sum_{i \in
{\cal I}} \eta_i \widetilde \lambda_i  + \sum_{i \in\widetilde{\cal I}}  \lambda_i \widetilde \eta_i &=&0,
\label{eq:momcons} \eea
which should be considered as four constraints on  $\eta_i$ and $\widetilde \eta_i$ which are easily satisfied provided the number of reference spinors is sufficient.  

\subsection{Factorization}

Next, consider a factorization channel of a subset of particles $\cal F$.  The complex deformation of the momenta in \Eq{eq:genshift} sends
\bea
P \rightarrow P(z) = P + z Q,
\label{eq:sumshift}
\eea
where $P$ is the original momentum flowing through the factorization channel and $Q$ is the net momentum shift, so
\bea
P=\sum\limits_{i\in {\cal F}} \lambda_i\widetilde\lambda_i ,&\quad&
Q =   \sum_{i \in
{\cal F}_\lambda} \eta_i \widetilde \lambda_i+ \sum_{i \in {\cal F}_{\widetilde\lambda}}  \lambda_i \widetilde \eta_i ,
\label{eq:defineQ}
\eea
where ${\cal F}_\lambda$ and ${\cal F}_{\widetilde \lambda}$ are intersection of ${\cal F}$ with ${\cal I}$ and $\widetilde{\cal I}$.  

As we will see, the physics depends crucially on whether $Q^2$ vanishes for all factorization channels or not.  First of all, the large $z$ behavior is affected because propagators in the complexified amplitude scale as
\bea
\frac{1}{(P+zQ)^2} &=& \left\{ 
\begin{array}{cc}
z^{-1} &,\quad Q^2=0\\
z^{-2} &,\quad Q^2\neq0
\end{array}
\right. ,
\eea
for a given factorization channel.  Second, there is a very important difference in the structure of the recursion relation depending on whether $Q^2$ vanishes in all channels.  If so, then each factorization channel has a simple pole at
\bea
z_* &=& -P^2 /2 P\cdot Q,
\eea
and the on-shell recursion relation takes the usual form,
\bea
{\cal M}(0) &=& \sum_{\cal F} \, \frac{1}{P^2} {\cal M}_{\cal F}(z_*) {\cal M}_{\bar{\cal F}}(z_*) + \textrm{(pole at $z=\infty$)},
\label{eq:RR1}
\eea
where the sum is over all factorization channels and intermediate states, and $ {\cal M}_{\cal F}$ and $ {\cal M}_{\bar{\cal F}}$ are on-shell amplitudes corresponding to each side of the factorization channel.  However, if $Q^2$ does not vanish, then each propagator is a quadratic in $z$ and thus carries conjugate poles at
\bea
z_{\pm} &=& \frac{ - P\cdot Q \pm \sqrt{(P\cdot Q)^2 - P^2 Q^2}}{Q^2}.
\eea
Summing over both of these roots, we find a new recursion relation,
\bea
{\cal M}(0) &=& \sum_{\cal F} \, \frac{1}{P^2} \left[ \frac{z_+ {\cal M}_{{\cal F}}(z_-)  {\cal M}_{\bar{\cal F}}(z_-)-z_{-} {\cal M}_{\cal F}(z_+)  {\cal M}_{\bar{\cal F}}(z_+)}{z_+ - z_-} \right]+ \textrm{(pole at $z=\infty$)}.
\label{eq:RR2}
\eea
Under conjugation of the roots, $z_+ \leftrightarrow z_-$, the summand is symmetric, so crucially, square roots always cancel in the final expression in the recursion relation.  Of course, the intermediate steps in the recursion are nevertheless quite cumbersome in this case.  

\subsection{Recursion Relations}

All known recursion relations can be constructed by imposing additional constraints on the momentum shift in \Eq{eq:genshift} beyond the condition of momentum conservation in \Eq{eq:momcons}.  In the absence of extra constraints, the reference spinors $\eta_i$ and $\widetilde\eta_i$ are arbitrary so by \Eq{eq:defineQ}, $Q^2\neq0$ generically.  In this case the recursion relation will have square roots in intermediate steps.  

On the other hand, if $Q^2 =0$, then $Q$ must factorized into the product of two spinors.  
If $Q$ is factorizable, then in the summand of \Eq{eq:defineQ} either the $\eta_i$ and $\lambda_i$ are proportional
or the $\widetilde \eta_i$ and $\widetilde\lambda_i$ are proportional.
For general external kinematics, {\it i.e.}~the $\lambda_i$ and $\widetilde\lambda_i$ are independent,  these proportionality conditions can involve at most one external spinor. As we will see, this implies two distinct classes of recursion relation which can accommodate $Q^2=0$. 

The first possibility is to shift only holomorphic spinors or only anti-holomorphic spinors subject to the constraint that the $\eta_i = c_i \eta$ and $\widetilde\eta_i = \widetilde c_i \widetilde \eta$ are all proportional to universal reference spinors $\eta$ and $\widetilde\eta$.  In each case, \Eq{eq:defineQ} factorizes into the form $Q=\eta(\ldots)$ and $Q=(\ldots)\widetilde\eta$, respectively. In mathematical terms, these scenarios correspond to the  $[0,m\rangle$-line and $[m,0\rangle$-line shifts,
\bea
\left[0,m   \right.\rangle\textrm{-line: }&& \left\{ 
\begin{array}{l}
 \lambda_i \rightarrow \lambda_i(z) =  \lambda_i +z c_i \eta, \quad i \in {\cal I}\nonumber \\
\sum_{i\in {\cal I}} c_i  \widetilde\lambda_i =0
 \end{array} \right. \\ \nonumber 
 \\
\left[m,0 \right. \rangle\textrm{-line: } &&\left\{
\begin{array}{l}
 \widetilde\lambda_i \rightarrow \widetilde\lambda_i(z) =  \widetilde\lambda_i + z\widetilde c_i \widetilde \eta, \quad i \in\widetilde{\cal I}\\
 \sum\nolimits_{i\in\widetilde{\cal I}} \widetilde{c}_i  \lambda_i =0
 \end{array}
 \right.  
 \label{eq:mline}
\eea
where the constraints on $c_i$ and $\widetilde c_i$ arise from momentum conservation.
Of course, the $[0,m\rangle$-line and $[m,0\rangle$-line shifts are simply  generalizations of the Risager shift with the only difference that here $m\leq n$ is arbitrary.

The second possibility is to shift only holomorphic spinors except for one or only anti-holomorphic spinors except for one.  In this case the reference spinors must be proportional to a spinor of a specific external leg, which we denote here by $\lambda_j$ or $\widetilde\lambda_j$.  Thus, in each case, $\eta_i = c_i \lambda_j$ and $\widetilde \eta_i = \widetilde c_i \widetilde \lambda_j$, so we again have factorization, but of the form $Q=\lambda_j(\ldots)$ and $Q=(\ldots)\widetilde\lambda_j$.
These correspond to $[1,m-1 \rangle$-line and $[m-1,1\rangle$-line shifts,
\bea
\left[1,m-1 \right. \rangle \textrm{-line: } && \left\{\begin{array}{ll} \lambda_i \rightarrow \lambda_i(z) =  \lambda_i + zc_i \lambda_j, &\quad i \in {\cal I} \nonumber \\
\widetilde\lambda_j \rightarrow \widetilde\lambda_j(z) =  \widetilde\lambda_j - z \sum_{ i \in {\cal I}}  c_i \widetilde \lambda_i, &\quad j =\widetilde{\cal I}\end{array} \right. \\ \nonumber \\
\left[m-1,1 \right. \rangle\textrm{-line: } && \left\{ \begin{array}{ll} \widetilde\lambda_i \rightarrow \widetilde\lambda_i(z) =  \widetilde\lambda_i + z \widetilde c_i \widetilde \lambda_j, & \quad i \in\widetilde{\cal I} \\
\lambda_j \rightarrow \lambda_j(z) =  \lambda_j -z \sum_{i \in\widetilde{\cal I}} \widetilde c_i \lambda_i, &\quad j = {\cal I}
\end{array} \right. 
 \label{eq:m1line}
\eea
where we have chosen a form such that momentum conservation is automatically satisfied.  Note that the case $m=2$ corresponds precisely to BCFW, so these shifts are a generalization of BCFW to arbitrary $m\leq n$.

Note that for $m\leq 3$, any momentum shift is necessarily of the form of the first or second possibility, so $Q^2=0$ automatically.  Thus, $Q^2\neq 0$ is only possible if $m>3$.

Remarkably, while the recursion relations in \Eq{eq:mline} and \Eq{eq:m1line} are naturally the generalizations of Risager and BCFW, they actually overlap for a specific choice of reference variables!  In particular, consider the $[0,m\rangle$-line and $[m,0\rangle$-line shifts in \Eq{eq:mline} for the case of $\eta = \lambda_j$ and $\widetilde \eta = \widetilde \lambda_j$, and modifying the constraint from momentum conservation such that $\sum_{i\in {\cal I}} c_i  \widetilde\lambda_i =\widetilde\lambda_j$ and $ \sum_{i\in\widetilde{\cal I}} \widetilde{c}_i  \lambda_i =\lambda_j$, respectively. In this case the recursion coincides with the form of the $[1,m-1\rangle$-line and $[m-1,1\rangle$-line shifts in \Eq{eq:m1line}, with a curious feature that $\lambda_j(z) = \lambda_j(1-z)$ and $\widetilde\lambda_j(z) = \widetilde\lambda_j(1-z)$.  We dub these ``soft'' shifts for the simple reason that when $z=1$ the amplitude approaches a soft limit. For $m=3$, the soft shift takes a particularly elegant form,
\bea
\textrm{3-line soft shift:} && \left\{\begin{array}{l}
 \lambda_1 \rightarrow \lambda_1(z) =  \lambda_1 + z \frac{[23]}{[21]} \lambda_3\\
  \lambda_2 \rightarrow \lambda_2(z) =  \lambda_2 + z \frac{[13]}{[12]} \lambda_3\\
    \widetilde\lambda_3 \rightarrow \widetilde\lambda_3(z) =  \widetilde\lambda_3 (1-z)
 \end{array} \right.\,\,\,\mbox{or}\,\,\,\,\left\{\begin{array}{l}
 \widetilde\lambda_1 \rightarrow \widetilde\lambda_1(z) =  \widetilde\lambda_1 + z \frac{\langle 23\rangle}{\langle 21\rangle} \widetilde\lambda_3\\
  \widetilde\lambda_2 \rightarrow \widetilde\lambda_2(z) =  \widetilde\lambda_2 + z \frac{\langle13\rangle}{\langle 12\rangle} \widetilde\lambda_3\\
    \lambda_3 \rightarrow \lambda_3(z) =  \lambda_3 (1-z)
 \end{array} \right.  .
 \label{eq:softshift}
\eea
This shift offers an on-shell prescription for taking a soft limit. We will not make use of this shift in this paper but leave a more thorough analysis of this soft shift for future work \cite{Cheung:20xx}.

\section{Large $z$ Behavior of Amplitudes}
\label{sec:z}

The recursion relations in \Eq{eq:RR1} and \Eq{eq:RR2} apply when the amplitude does not have a pole at $z=\infty$.  In this section we determine the conditions under which this boundary term vanishes. Although one could study the boundary term in BCFW or Risager shift instead, as in \Ref{Feng:2009ei,Jin:2014qya}, we will not proceed in this direction. Concretely, take the $n$-point amplitude, $\cal M$, deformed by an $m$-line shift where $m\leq n$.  At large $z$, the shifted amplitude describes the physical scattering of $m$ hard particles in a soft background parametrizing the remaining $n-m$ external legs.  Thus, we can determine the large $z$ behavior by applying a background field method: we expand the original Lagrangian in terms of soft backgrounds and hard propagating fluctuations, then compute the on-shell $m$-point ``skeleton'' amplitude, $\widetilde{M}$.  If the skeleton amplitude vanishes at large $z$, then the boundary term is absent and the recursion relation applies.   A similar approach was applied in \Ref{ArkaniHamed:2008yf} for BCFW for the case of a hard particle propagator, {\it i.e.}~the skeleton amplitude for $m=2$.  

Crucially, it will not be necessary to explicitly compute the skeleton amplitude.  Rather, from Lorentz invariance, dimensional analysis, and the assumption of local poles, we will derive general formulae for the large $z$ behavior of $m$-line shifts of $n$-point amplitudes.  Hence, our calculation of the large $z$ scaling combines and generalizes two existing proofs in the literature relating to the BCFW~\cite{ArkaniHamed:2008yf} and all-line recursion relations\cite{Cohen:2010mi}.

\subsection{Ansatz}

The basis of our calculation is a general ansatz for the $m$-point skeleton
amplitude for $m \leq n$,
\begin{equation}
\widetilde{{\cal M}}=\widetilde{g} \times \sum_{\rm diagrams} \left( F  \times  \prod_{\rm vectors}   \varepsilon  \times   \prod_{\rm fermions}   u \right)
\label{eq:ansatz}
\end{equation}
where the sum is over Feynman diagrams $F$, which are contracted into products over the polarization vectors $\varepsilon$ and fermion wavefunctions $u$ of the hard particles\footnote{Note that polarization vectors arise from any particle of spin greater than or equal to one.}. Here $\widetilde{g}=g\times B$ where $g$ is a product of Lagrangian coupling constants and $B$ is a product of soft field backgrounds and their derivatives.  Note that $\widetilde{g}$ has free Lorentz indices since it contains insertions of the soft background fields and their derivatives. Crucially, since $B$ is comprised of backgrounds, it is always non-negative in dimension, so $[B]\geq 0$ and
\begin{equation}
[\widetilde{g}]=[g]+[B]\ge [g].
\end{equation}
For the special case of gravitational interactions, each insertion of the background graviton field is accompanied by an additional coupling suppression of by the Planck mass, so $[\widetilde{g}]=[g]$.  This is reasonable because the background metric is naturally dimensionless so insertions of it do not change the dimensions of the overall coupling.


Note the skeleton amplitude receives dimensionful contributions from every term in \Eq{eq:ansatz} except the vector polarizations, so
\begin{equation}
[\widetilde{{\cal M}}]=4-m=[\widetilde{g}]+[F]+ \sum_{\rm fermions}1/2,
	\label{eq:M_dim}
\end{equation}
via dimensional analysis.  This fact will be crucial for our calculation of the large $z$ scaling of the skeleton amplitude for various momentum shifts and theories.


\subsection{Large $z$ Behavior}
We analyze the large $z$ behavior of \Eq{eq:ansatz}. The contribution from each Feynman diagram $F$ can be expressed as a ratio of polynomials in momenta, so $F=N/D$. Here $N$ arises from interactions while $D$ arises from propagators.  We define the large $z$ behavior of the numerator and denominator as $\gamma_N$ and $\gamma_D$ where
\bea
N \sim z^{\gamma_N} , &\quad&
D \sim z^{\gamma_D}.
\label{eq:gammadefs}
\eea
We now compute the large $z$ behavior of the external wavefunctions, followed by that of the Feynman diagram numerator and denominator, and finally the full amplitude.

\paragraph{External Wavefunctions.}
First, we study the contributions from external polarization vectors and fermion wavefunctions.  For convenience, we define a ``weighted'' spin, $\widetilde s$, for each shifted leg of $+/-$ helicity, which is simply the spin $s$ multiplied by $+$ if the angle/square bracket is shifted and $-$ if the square/angle bracket is shifted.  In mathematical terms,
\bea
	\widetilde s &=& s\times \left\{ \begin{array}{ll}
	+  ,&\quad 
	\textrm{good shift}\\
	- , &\quad  
	\textrm{bad shift}
	\end{array}
	\right. ,
\eea
where good and bad shifts denote the correlation between helicity and the shift of spinor indicated in \Eq{eq:goodbad}.  As we will see, a multiplier of $+$/$-$ tends to improve/worsen the large $z$ behavior.   In terms of the weighted spin, it is now straightforward to determine how the large $z$ scaling of the polarization vectors and fermion wavefunctions,
\bea
	\text{external wavefunction} &\sim& \left\{ 
	\begin{array}{ll}
		z^{-\widetilde{s}},&\quad \text{boson} \\
		z^{-(\widetilde{s}-1/2)},&\quad \text{fermion}
	\end{array}
	\right. .
	\label{eq:escaling}
\eea
so more positive values of $\widetilde s$, corresponding to good shifts, imply better large $z$ convergence.

\paragraph{Numerator and Denominator.}

\begin{figure}[t]
	\centering
	\vspace*{-1.75cm}
	\includegraphics[trim= 2cm 0mm 16cm 12cm, clip, scale=1.2]{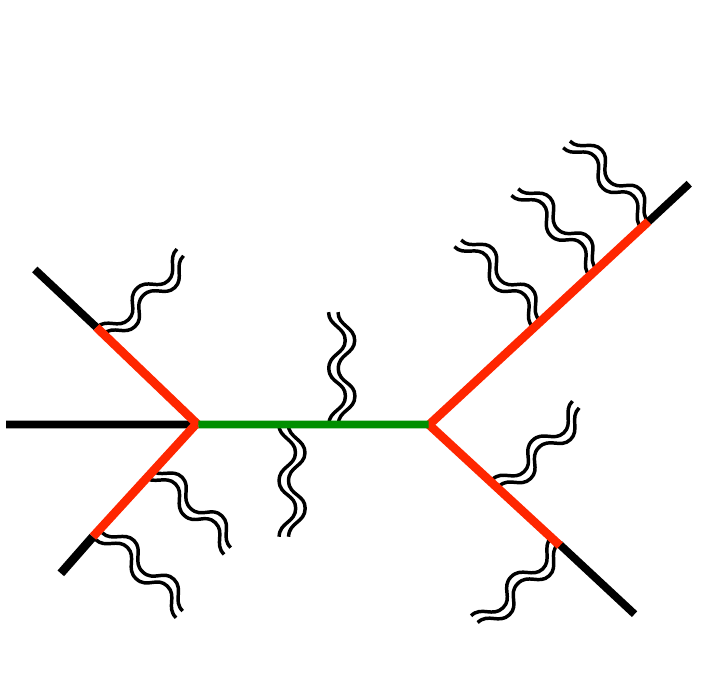}
	\caption{A skeleton diagram for an $Q^2 \neq 0$ shift. Here straight lines are hard particles and curved lines are soft backgrounds. Color segments are propagators, and red and green denotes those that scale as $1/z$ and $1/z^2$ at large $z$, respectively.}
	\label{fig:problem_diagram}
\end{figure}

The numerator $N$ of each Feynman diagram depends sensitively on the dynamics. However, for a generic shift, we can conservatively assume no cancellation in large $z$ so the numerator scales at most as its own mass dimension, 
\bea
\gamma_N \leq  [N].
\label{eq:numeratorz}
\eea 
The denominator $D$ comes from propagators which are fully dictated by the topology of the diagram. Each propagator can scale as $1/z^2$ or $1/z$ at large $z$, depending on the details of shifts. Thus, the large $z$ behavior of denominator is constrained to be within
\be
	\frac{[D]}{2} \leq \gamma_{D} \leq {[D]}.
	\label{eq:denominator}
\ee
For the $Q^2=0$ shifts, every propagator scales as $1/z$ so $\gamma_D ={[D]/2}$. On the other hand, for the $Q^2\neq 0$ shifts, we would naively expect that there is a $1/z^{2}$ from each propagator given that the reference spinors are arbitrary. However, this reasoning is flawed due to an important caveat.  Since the theory contains soft backgrounds, the Feynman diagram can have 2-point interactions of the hard particle induced by an insertion of the soft background. If the 2-point interactions occur before the hard particle interacts with another hard particle, then $Q$ is simply the momentum shift of a single external leg, so $Q^2=0$ accidentally, and the corresponding propagator scales as $1/z$ rather than $1/z^2$.  It is simple to see that the number of such propagators is $[D]-\gamma_D$.    See \Fig{fig:problem_diagram} for an illustration of this effect. Thus the large $z$ behavior is constrained within the range of \Eq{eq:denominator}.

From our knowledge of Feynman diagrams, we can further relate the total number of propagators to the number of hard external legs, $m$, and the valency of the interactions, $v$, yielding
\begin{equation}
\frac{[D]}{2} \leq \left( \frac{m - v}{v-2}\right) + [B],
\label{eq:propagator}
\end{equation}
where $v\geq 3$ is the valency of the interaction vertices in the fundamental theory and  the $[B]$ term arises because we have conservatively assumed that every single background field insertion contributes to a 2-point interaction to the amplitude.

\paragraph{Full Amplitude.} Combing in the large $z$ scaling of the external wavefunctions in \Eq{eq:escaling}  with that of the numerator and denominator of the  the Feynman diagram in \Eq{eq:gammadefs}, we obtain
\bea
	\gamma &=& \gamma_N -\gamma_D -\sum_{\text{bosons}}\widetilde{s}-\sum_{\text{fermions}}(\widetilde{s}-1/2) 
	\nonumber\\
	&\leq& 4-m-[g]-\sum_{\text{all}}\widetilde{s}+[D]-\gamma_D-[B],
	\label{eq:amp_bound1}
\eea
where in the second line we have plugged in the inequality from  \Eq{eq:numeratorz}, replaced $[N] = [F] + [D]$, and eliminated $[F]$ by solving \Eq{eq:M_dim}. This is the master formula from which we will derive corresponding large $z$ behaviors in $Q^2\neq 0$ and $Q^2=0$ shifts.
As expected, the above bound can be improved for $Q^2=0$ shifts because in this case the product of any two hard momenta only scales as $z$ rather than $z^2$. We render the specific derivation in subsequent sections.

The general formula in \Eq{eq:amp_bound1} can be reduced to more illuminating forms by making the assumption of specific shifts. We consider the large $z$ behavior for the $Q^2 \neq 0$  and $Q^2=0$ shifts in turn.

\subsubsection{($Q^2 \neq0$)}
To start, we calculate the large $z$ behavior for a general momentum shift defined in \Eq{eq:genshift}. As noted earlier, for arbitrary reference spinors, $Q^2\neq 0$ as long as $m\geq 3$, which we assume here. The large $z$ behavior is given by \Eq{eq:amp_bound1}. The offset $[D]-\gamma_D$ is the number of propagators with $Q^2=0$ as discussed before.  As shown for an example topology in \Fig{fig:problem_diagram}, there is at least one  soft background associated with each propagator for which $Q^2=0$. The canonical dimensions of fields leads to $[D]-\gamma_D-[B]\leq 0$. We conclude that
\bea
\boxed {\gamma \leq  4-m-[g]-\sum_{\rm all}\widetilde{s}.}
	\label{eq:bestbound}
\eea
The large $z$ convergence is best for the largest possible value for $\widetilde{s}$, which occurs if we only apply good shifts to external legs, so $\widetilde{s} =s$.
As we will see, this particular choice has the best large $z$ behavior of any shfit. There is an inherent connection between $Q^2\neq0$ and improved $z$ behavior of the amplitude, simply because in this case, propagators fall off with $z^2$ in diagrams.

\subsubsection{($Q^2 =0$)}

Next, we compute the large $z$ behavior of the momentum shift in \Eq{eq:genshift} when $Q^2=0$. In these shifts, substituting $\gamma_D=[D]/2$ and \Eq{eq:propagator} into \Eq{eq:amp_bound1} yields
\begin{equation}
		\boxed{\gamma 
		\leq 1 - \left(\frac{v-3}{v-2}\right)(m-2) - [g]-\sum_{\rm all} \widetilde s.}
	\label{eq:bound1}
\end{equation}
For trivalent interactions, $v=3$,  the bound is independent of $m$.  For quadrivalent vertices, $v=4$, the bound improves for larger numbers of shifted legs, $m$.

We showed previously that $Q^2=0$ can only occur for the $[0,m\rangle$-, $[m,0\rangle$-, $[1,m-1\rangle$-, and $[m-1,1\rangle$-line shifts defined in \Eq{eq:mline} and \Eq{eq:m1line}.  Hence, we can learn more by considering the specific form of the large $z$ shifts. In the subsequent sections we consider each of these cases in turn to derive additional bounds on the large $z$ behavior.

\paragraph{$[0,m\rangle $-Line and $[m,0\rangle $-Line Shifts.}
The $[0,m\rangle $-line and $[m,0\rangle $-line shifts defined in \Eq{eq:mline} are a generalization of the Risager momentum shift, for which $Q^2=0$.  To begin, let us consider the large $z$ behavior of the $[0,m\rangle $-line shift; an identical argument will of course hold for the $[m,0\rangle $-line shift.  We only have to keep track of holomorphic spinors, since anti-holomorphic spinors are not shifted.   To conservatively bound the large $z$ behavior of the numerator of \Eq{eq:ansatz}, we can simply sum the total number of holomorphic spinors and divide by two, since the reference spinors are proportional and thus vanish when dotted into each other.  However, note that we must remember to count the holomorphic spinors coming from the numerator $N$ as well as from the soft background $B$ and external wavefunctions.  Overall \Eq{eq:escaling} gives the correct number of holomorphic spinors. Including all contributions yields
\bea
\gamma &\leq&
\frac{1}{2}\left( [N]+n_{B}-\sum_{\text{bosons}}\widetilde{s}-\sum_{\text{fermions}}(\widetilde{s}-1/2)-[D] \right),
\label{eq:amp_bound2}
\eea
where $n_B$ is the number of holomorphic spinors indices that come from soft background insertions.
Again solving for $[F]$ with \Eq{eq:M_dim}, and applying our arguments to both shifts, the large $z$ behavior is
\bea
\boxed{
\gamma \leq \left\{
\begin{array}{ll} 
\frac{1}{2} \left( 4-m-[g]-\sum\limits_{\rm all} h +\Delta\right), & \quad [0,m\rangle \textrm{-line} \quad \\
\frac{1}{2} \left( 4-m-[g]+\sum\limits_{\rm all} h +\Delta\right), & \quad [m,0\rangle \textrm{-line}
\end{array}
\right.} 
\label{eq:bound_h}
\eea
where $h$ denotes helicity and we have defined
\bea
\Delta=n_B-[B].
\label{eq:Deltadef}
\eea
In a theory with only spin $s\leq 1$ fields, soft background insertions contribute at most one holomorphic or anti-holomorphic spinor index to be contracted with. Thus, $n_B$ is balanced by the dimension $[B]$, so $\Delta\leq 0$ in these theories. On the other hand, for a theory with spin $s\leq 2$ fields, {\it e.g.,} gravitons, then an insertion of a graviton background yields two spinor indices but only with one power of mass dimension.   For these two cases we thus find
\bea
	\Delta &\leq& \left\{
	\begin{array}{ll} 
		0, & \quad \text{theories with } s\leq 1 \\
		n-m, & \quad \text{theories with }s\leq 2 \\
	\end{array}
	\right. .
\label{eq:boundh_sup} 
\eea
\Eqs{eq:bound_h}{eq:boundh_sup} together give our final answer.   For an all-line shift, $m=n$, so $\Delta=0$ and this bound reduces to known result from~\Ref{Cohen:2010mi}. Note that in some cases \Eq{eq:bound1} is stronger than \Eq{eq:bound_h} so we have to consider both bounds at the same time.

\paragraph{$[1,m-1\rangle $-Line and $[m-1,1\rangle $-Line Shifts.}
The $[1,m-1\rangle $-line and $[m-1,1\rangle $-line shifts defined in \Eq{eq:m1line} are a generalization of the BCFW momentum shift, for which $Q^2=0$.  
To start, consider a $[1,m-1\rangle $-line shift, where particle $j$ has a shifted in anti-holomorphic spinor and all other shifts are on holomorphic spinors. To determine the large $z$ behavior of the $[1,m-1\rangle$-line shift, we start with our earlier result on the $[0,m\rangle $-line shift. By switching the deformation on particle $j$ from a shift of $|j]$ to a shift of $|j\rangle$, all the angle brackets associated with $j$ changes their scaling from $1$ to $z$ at large $z$ for generic choice of $\widetilde{c}_i$ in \Eq{eq:m1line}. In the mean time, all square brackets involving particle $j$ reduce from $z$ to $1$ because the reference spinor is $|j]$. The change in large $z$ behavior from a $[0,m\rangle $-line shift to a $[1,m-1\rangle $-line shift is exactly the difference of the degrees between anti-holomorphic and holomorphic spinors of $j$, which is fixed by little group.
Applying the reasoning to both shifts, we obtain
\bea
\boxed{
\gamma \leq \left\{
\begin{array}{ll} 
\frac{1}{2} \left( 4-m-[g]-\sum\limits_{\rm all} h +\Delta\right)+2h_j, & \quad [1,m-1\rangle \textrm{-line} \quad \\
\frac{1}{2} \left( 4-m-[g]+\sum\limits_{\rm all} h +\Delta\right)-2h_j, & \quad [m-1,1\rangle \textrm{-line}
\end{array}
\right.} 
\label{eq:bound_mixed}
\eea
where $h_j$ is the helicity of particle $j$.
We then see that the $[1,m-1\rangle$-line shift improves large $z$ behavior of $[0,m\rangle$-line shift if $h_j>0$.

The above argument has a caveat in the special case of the $[1,1\rangle$-line shift, {\it i.e.} the BCFW shift. Shifting the anti-holomorphic spinor of particle $i$ and the holomorphic spinor of particle $j$, then the angle bracket $\ab{ij}$ does not scale as $z$ at large $z$ so \Eq{eq:bound_mixed} does not apply. Nevertheless, we can still use \Eq{eq:bound1} which is valid for BCFW shift.

\section{On-Shell Constructible Theories}
\label{sec:th}
In this section we at last address the question posed in the introduction: \emph{what is the simplest recursion relation that constructs all on-shell tree amplitudes in a given theory?}   To find an answer we consider the $Q^2\neq 0 $ momentum shift defined in \Eq{eq:genshift} and the $Q^2=0$ momentum shifts defined in \Eq{eq:mline} and \Eq{eq:m1line}.  We utilize our results for the large $z$ behavior in \Eq{eq:bestbound}, \Eq{eq:bound1},
\Eq{eq:bound_h}, and \Eq{eq:boundh_sup}.
Throughout the rest of the paper we restrict to the good momentum shifts defined in \Eq{eq:goodbad}.  Thus, we only shift the holomorphic spinors of plus helicity particles and the anti-holomorphic spinors of negative helicity particles, and the weighted spin of each leg is equal to its spin, $\tilde s = s$. Unless otherwise noted, we henceforth denote any scalar/fermion/gauge boson/graviton by $\phi/\psi/A/G$.

\subsection{Renormalizable Theories}
To begin we consider the generic momentum shift defined in \Eq{eq:genshift}, which has large $z$ behavior derived in \Eq{eq:bestbound}.  Since a renormalizable theory only has marginal and relevant interactions,  the mass dimension of the product of couplings in any scattering amplitude is $[g]\geq 0$.  Plugging this into  \Eq{eq:bestbound}, we find that a 5-line shift suffices to construct any amplitude.  This is also true for the 5-line shifts defined in \Eq{eq:mline} and \Eq{eq:m1line}, whose large $z$ scaling is shown in \Eq{eq:bound_h} and \Eq{eq:bound_mixed} by conservatively plugging in $\Delta=0$ for renormalizable theories.
Consequently, 5-line recursion relations provide a purely on-shell, tree-level definition of any renormalizable quantum field theory.  We must take as input the three and four point on-shell tree amplitudes, but this is quite reasonable, as a renormalizable Lagrangian is itself specified by interactions comprised of three or four fields.

Fortunately, simpler recursion relations are sufficient to construct a more restricted but still enormous class of renormalizable theories.  
To see this, consider a general 3-line momentum shift and its associated large $z$ behavior shown in \Eq{eq:bound1}.  The amplitude vanishes at large $z$   provided the sum of the spins of the three shifted legs is greater than one.  This is automatic if all three shifted particles are vectors or fermions.  Such a shift can always be chosen unless the amplitude is composed of {\it i}) one vector and scalars, {\it ii}) two fermions and scalars, or {\it iii}) all scalars.  In case {\it i}), we can apply a 3-line shift of the form   $[\lrbraces{\phi,\phi},\lrbraces{A^+}\rangle$ or $[\lrbraces{A^-},\lrbraces{\phi,\phi}\rangle$, while in case {\it ii}), we can apply a 3-line shift of the form $[\lrbraces{\phi,\phi},\lrbraces{\psi^+}\rangle$ or $[\lrbraces{\psi^-},\lrbraces{\phi,\phi}\rangle$.  In both cases the large $z$ behavior is vanishing according to \Eq{eq:bound_mixed}.  Hence, any amplitude with an external vector or fermion is 3-line constructible.   

This leaves case ${\it iii})$, which is the trickiest scenario: an amplitude with only external scalars.  In general, such an amplitude is not 3-line constructible, but the story changes considerably if the scalars are covariant under a global or gauge $U(1)$ symmetry. Concretely, consider a 3-line shift of the form $[\lrbraces{\phi,\phi,\phi},0\rangle$ or $[0,\lrbraces{\phi,\phi,\phi}\rangle$.  Moreover, let us assume that the shifted legs carry a net charge under the scalar $U(1)$ which is not equal to the charge of any other scalar in the spectrum.  In this case, invariance under the scalar $U(1)$ requires that the amplitude has more than one additional external scalar with unshifted momenta.  The charge cannot be accounted for by an external fermion with unshifted momenta, since the amplitude only has external scalars.
From the perspective of the skeleton diagram describing the scattering of three hard particles in a soft background, the additional scalars correspond to more than one insertions of a soft scalar background, so as defined in \Eq{eq:Deltadef}, $\Delta < -1$.  Thus, according to \Eq{eq:bound_h}, the 3-line shift has vanishing large $z$ behavior and the associated amplitudes are constructible.  Note that the charge condition we have assumed is automatically satisfied if every scalar in the theory has equal charge under the scalar $U(1)$ and we shift three same-signed scalars.

It seems impossible for this 3-line recursion to construct all  equal-charged $U(1)$ scalar amplitudes, especially with the presence of quartic potential. However, as three same-signed scalars only available from six points, this 3-line recursion still takes three and four point amplitudes as seeds. The information of quartic potential still enters to this special 3-line recursion. We will demonstrate with a simple $\phi^4$ theory in next section.

Putting everything together, we have shown that a 3-line shift can construct any amplitude with a vector or fermion, and any amplitude with only scalars if every scalar carries equal charge under a $U(1)$ symmetry.  Immediately, this implies that any theory of solely vectors and fermions---{\it i.e.} any gauge theory with arbitrary matter content---is constructible\footnote{Note that such theories are constructible from BCFW, via a shift of any vector~\cite{Cheung:2008dn} or any same helicity fermions~\cite{Schwinn:2007ee}.}.  Moreover, all amplitudes in Yukawa theory necessarily carry an external fermion, so these are likewise constructible.  The standard model is also 3-line constructible simply because it has a single scalar---the Higgs boson---which carries hypercharge.   Finally, we observe that all supersymmetric theories are constructible.  The reason is that without loss of generality, the superpotential for such a theory takes the form $W = \lambda_{ijk} \phi_i \phi_j \phi_k$, where we have shifted away Polonyi terms and eliminated quadratic terms to ensure a massless spectrum.  For such a potential there is a manifest $R$-symmetry under which every chiral superfield has charge $2/3$.  Consequently, all complex scalars in the theory have equal charge under the $R$-symmetry and all amplitudes are 3-line constructible.  This then applies to theories with extended supersymmetry as well.  The conditions for on-shell constructibility in some familiar theories is summarized in Tab.~\ref{table:renormalizable_summary}.

\subsection{Non-renormalizable Theories}
In what follows, we first discuss non-renormalizable theories which are constructible, {\it i.e.}~for which all amplitudes can be constructed.   As we will see, this is only feasible for a subset of non-renormalizable theories, so in general, the covering space of recursion relations does not provide an on-shell formulation of all possible theories.  Second, we consider scenarios in which some but not all amplitudes are constructible within a given non-renormalizable theory.  In many cases, amplitudes involving a finite number of higher dimension operator insertions can often be constructed by our methods.  

Our analysis will depend sensitively on the dimensionality of coupling constants, which we saw earlier have a huge influence on the the large $z$ behavior under momentum shifts. Table~\ref{table:coupling_summary} summarizes the dimensions of coupling constants in various theories\footnote{As pointed out in \Ref{Cohen:2010mi}, we need to choose the highest dimension coupling if there are multiple of coupling constants.}. 
Here $v$ is the (minimal) valency of the vertex. $F$ and $R$ is defined as vector field strength and Riemann tensor, respectively, and we have omitted indices and complex conjugations for simplicity. The superscript of an external state specifies its helicity.
We keep the number of operator insertions, $u$, as a free parameter. At tree-level, it is constrained by  by the number of propagators, $u\leq[D]/2+1$, where $[D]/2$ is given in \Eq{eq:propagator}.

\paragraph{Constructible Theories.}

\begin{table}[t]
	\centering
	\begin{tabular}{ c ||c | c |c| c| c|}

		Theory  & $\phi^v$ & $\psi^v$ & $F^v$  & $R^v$ & Einstein (+ Maxwell) \\
		\hline \hline
		$[g]$ & $u(4-v)$ & $u(4-3v/2)$ & $u(4-2v)$  & $2-n-2u(v-1)$ & $2-n$
	\end{tabular}
	\caption{The dimensionality of the coupling constant, $[g]$, for an $n$-point amplitude, where $u$ denotes the number interaction vertices, which have minimal valency $v$.}
	\label{table:coupling_summary}
\end{table}

To start, consider a theory of scalars interacting via a $\phi^v$ operator. Following \Eq{eq:propagator}, and using that the dimensionality of backgrounds is positive, $[B] \geq 0$, we can bound the number of propagators by $[D]/2\geq (m-v)/(v-2)$ for in a $m$-point skeleton amplitude. The number of interaction vertices exceeds the number of propagators by one, so $u=[D]/2+1$. In a $[m,0\rangle$-line shift, substituting $[g]=u(4-v)$ from Table~\ref{table:coupling_summary}, and plugging into \Eq{eq:bound_h} with  $\Delta= -[B] \leq 0$ for scalars, we have
\bea
	\gamma &\leq& \,\frac{v-m}{v-2}.
	\label{eq:gammascalar}
\eea
Thus, we find that all amplitudes in $\phi^v$ theory are constructable for an $[m,0\rangle$-line shift where $m > v$ and the $v$ point amplitude is taken as the input of the recursion relation\footnote{In fact, $m=v$ suffices to construct any amplitude with $v+1$ points or above. This can be derived if we treat soft background in $[D]/2$ more carefully.}.  Since the scalars have no spin, this large $z$ also applies for the conjugate $[0,m\rangle$-line shift.  Of course, this conclusion is completely obvious from the perspective of Feynman diagrams. In particular, since $\phi^v$ theory does not have any kinematic numerators, its amplitudes are constructible provided there is even one hard propagator, which happens as long as $m>v$.

Analogously, consider a theory of fermions interacting via $\psi^v$ operators.  Conservatively, we assume all soft fermions in the skeleton amplitude are emitted from $Q^2=0$ propagators
\begin{equation}
[D]-\gamma_D-[B]=\frac{n_f}{v-2}-\frac{3}{2}n_f,
\end{equation}
where $n_f$ is the number of soft fermion insertions.
Substituting the above equation and the number of vertices $u=(m+n_f-2)/(v-2)$ into the large $z$ behavior for a general $m$-line shift in \Eq{eq:amp_bound1}, we find exactly the same expression for $\gamma$ in \Eq{eq:gammascalar}.  Thus, all amplitudes in $\psi^v$ theory are constructible with generic $m$-line shift for $m > v$, and taking the $v$ point amplitude as an input. Again, it is not surprising from Feynman diagrams. Note that we here required a general $m$-line shift with $Q^2\neq0$, such that the fermionic propagators $\slashed{P}/P^2$ scale as $1/z$ at large $z$. On the other hand, the recursion relation cannot work for a $Q^2= 0$ momentum shift because the fermionic propagators do not fall off at large $z$.

It is straightforward to generalize the arguments above to a theory of scalars and fermions interacting via a $\phi^{v_1}\psi^{v_2}$.  We find that this theory is fully constructible with a general $m$-line shift for $m > v_1 +v_2$.

Finally we consider perhaps the most famous constructible non-renormalizable theory: gravity. As is well-known, all tree-level graviton scattering amplitudes can be recursed via BCFW \cite{ArkaniHamed:2008yf}, taking the 3-point amplitudes as input.    Still, let us see how each of our $m$-line shifts fare relative to BCFW.   Throughout, we consider only good shifts, as defined in \Eq{eq:goodbad}.  Using \Eq{eq:bestbound} and \Eq{eq:bound1}, the large $z$ behaviors of $m$-line shifts are
\bea
\gamma &\leq& \left\{
	\begin{array}{ll}
		n+2-3m,\quad & Q^2\neq 0\text{ shift}\\
		n-1-2m,\quad & Q^2= 0\text{ shift}
	\end{array}
	\right. .
\eea
With the $Q^2\neq 0$ shifts, we can always construct an $n$-point amplitude with $m>(n+2)/3$. Applying the above result to NMHV amplitudes for $m=3$, we find ${\cal M} \lesssim z^{n-7}$ under a Risager 3-line shift, consistent with the known behavior $z^{n-12}$~\cite{Bianchi:2008pu}. Generally, graviton amplitude can be constructed with $Q^2=0$ shifts if $m\geq n/2$. \Ref{Cohen:2010mi} shows amplitudes with total helicity $|h|\leq 2$ cannot be constructed from anti-holomorphic/holomorphic all-line shift. We see this can be resolved if we choose to do ``good'' shift on only plus or negative helicity gravitons.
Our large $z$ analysis predicts the scaling grows linearly with $n$ and this is indeed how the real amplitude behaves. From this point of view, the amplitude behaves surprisingly well under BCFW shift because the scaling doesn't grow as $n$ increases.

An interesting comparison of our large $z$ behavior is to use the KLT relations~\cite{Kawai:1985xq}. Consider the large $z$ behavior of $n$ point amplitudes under a $(m\geq4)$-line $Q^2\neq0$ shift. A $n$ point graviton amplitude ${\cal M}_{\rm grav}$ can be schematically written as a ``square'' of gauge amplitudes ${\cal M}^2_{\rm gauge}$ by the KLT relation
\bea
{\cal M}_{\rm grav}\big|_{z\rightarrow \infty} &\sim& s^{n-3}{\cal M}_{\rm gauge}^2\big|_{z\rightarrow \infty} \nonumber \\
z^{n+2-3m} & \geq & z^{n-3} z^{8-4m}=z^{n+5-4m},
\eea
where we neglect all the permutation in particles and details of $s$-variables\footnote{The inequality holds for $m\ge4$ which is satisfied in any $Q^2\neq 0$ shift.}. The KLT relation actually predicts a better large $z$ behavior than our dimensional analysis.

\paragraph{Constructible Amplitudes.}

The above non-renormalizable theories are some limited examples which can be entirely defined by our on-shell recursions. Modifying these theories generally breaks the constructibility! For instrance, a theory of higher dimensional operator $\partial^2\phi^{v}$ cannot be constructed. This is clear from Feynman diagrams because the derivatives in vertices compensate the large $z$ suppression from propagators. This implies the chiral Lagrangian is not constructible even with the best all-line shift\footnote{The chiral Lagrangian has the additional complication that there is an infinite tower of interactions generated at each order in the pion decay constant. To overcome this, it is important to use soft limits to relate them and construct the amplitudes~\cite{Kampf:2012fn}.}.
In gauge theories, we cannot construct amplitudes where all vertices are higher dimensional $F^v$ operators either.

Fortunately, we are usually interested in effective theories with some power counting on higher dimensional operators. If the number of operator insertions is fixed, then we can construct amplitudes with generic multiplicity. To illustrate this, consider amplitudes in a renormalizable theory (spin $\leq 1$) with a single insertion of a $d$-dimensional operator.
If we apply a general $m$-line $Q^2\neq 0$ momentum shift, \Eq{eq:bestbound} gives
\begin{equation}
\gamma_{\rm gen} \leq d-m-s.
\end{equation}
In the worst case scenario, $s=0$, we see an $(d+1)$-line shift suffices to construct any such amplitude. For $[0,m\rangle$- and $[m,0\rangle$-line shifts, the sum of their large $z$ scaling is
\begin{equation}
\gamma_{[0,m\rangle}+\gamma_{[m,0\rangle} \leq d-m,
\end{equation}
where we use $\Delta = 0$ for theories with spin $\leq 1$.
The amplitude can always be constructed from one of them provided $m> d$. We see the input for recursion relations are all amplitudes with $d$ points and below. It is not surprising. After all, we need this input for a $\phi^v$ operator. If the amplitude has higher total spin/helicity, less deformation is needed to construct it. We will demonstrate this with $F^v$ operator in next section. The result is similar to the conclusion of \Ref{Cohen:2010mi}, but we can be more economical by choosing $(d+1)$-line or less rather than an all-line shift.

\section{Examples}
\label{sec:ex}
In this section, we illustrate the power of our recursion relations in various theories. The calculation is straightforward once the large $z$ behavior is known.

\paragraph{YM + $\psi$  + $\phi$.}
Consider a gauge theory with fermion and scalar matter in the adjoint representation.  In addition to the gauge interactions, there are Yukawa interactions of the form $ \text{Tr}(\phi\lrbraces{\psi,\psi})$.
Here we construct the color-ordered amplitude ${\cal M}(\psi^-,\psi^-,\phi,\phi,\phi)$ via a 3-line shift $[\lrbraces{2},\lrbraces{3,4}\rangle$. The  seed amplitudes for the recursion relation are
\begin{equation}
	\begin{split}
	{\cal M}(\psi^-,\psi^-,\phi)&= y\ab{12}  \\
	{\cal M}(\psi^-,\psi^+,A^-)&=g\ab{31}^2/\ab{12} \\
	{\cal M}(\phi,\phi,A^-) &=g\ab{31}\ab{23}/\ab{12} \\
	{\cal M}(\phi,\phi,\phi,\phi)&=g^2\left( 1 + \frac{[13]^2[24]^2}{[12][23][34][41]}\right) \\
	{\cal M}(\psi^-,\psi^+,\phi,\phi)&=
	g^2 \frac{[23][24]}{[12][34]}-y^2\frac{[24]}{[41]},
	\end{split}
\end{equation}
where $y$ and $g$ are the Yukawa and gauge coupling constants, respectively.
There are only two non-vanishing factorization channels. Based on these seeds, it's straightforward to write down
\begin{equation}
	\begin{split}
		{\cal M}(\psi^-,\psi^-,\phi,\phi,\phi) &=
		yg^2\left(\frac{1}{[12]}+\frac{[14]^2[35]^2}{[13][12][34][45][51]}-\frac{[35][34]}{[13][23][45]}\right)+y^3\frac{[35]}{[23][51]}.
	\end{split}
\end{equation}
Note that the spurious pole $[13]$ cancels between terms. From the final answer, we see that neither the BCFW shifts, like $[\lrbraces{2},\lrbraces{3}\rangle$ and $[\lrbraces{1},\lrbraces{2}\rangle$, nor the Risager shift on $[\lrbraces{2,3,4},0\rangle$ can construct the amplitude.  Thus, a 3-line shift such as $[\lrbraces{2},\lrbraces{3,4}\rangle$ is necessary to construct theories with both gauge and Yukawa interactions.

	

\paragraph{$\mathcal{N}=1$ SUSY.}
We have shown all massless supersymmetric theories are 3-line constructible. Consider a $\mathcal{N}=1$ supersymmetric gauge theory with an $SU(3)$ flavor multiplet of adjoint chiral multiplets $\Phi_{a}$.  We assume a superpotential
\begin{equation}
W=i\lambda \text{Tr}(\Phi_a[\Phi_b,\Phi_c]),
\end{equation}
where $a,b,c$ are fixed $SU(3)$ flavor indices, no summation implied.
We apply our recursion relations on the (color-ordered) 6-point scalar amplitude ${\cal M}(\phi^-_a,\phi^-_b,\phi^-_c,\phi^+_c,\phi^+_b,\phi^+_a)$, where the superscripts and subscripts denote $R$-symmetry and flavor indices, respectively.  In the massless limit, all scalars in the chiral multiplets carry equal $R$-charge. Therefore we can shift the three holomorphic scalars, namely, $[\lrbraces{1,2,3},0\rangle$.
The relevant lower point amplitudes for recursion are
\begin{equation}
\begin{split}
{\cal M}(A^-,\phi^{\pm}_a,\phi^{\mp}_a)&=
\frac{\ab{31}\ab{12}}{\ab{23}} \\
{\cal M}(\phi^{-}_a,\phi^{-}_b,\phi^{+}_b,\phi^{+}_a)&=
\frac{\ab{13}\ab{42}}{\ab{41}\ab{23}}+(1-\lambda^2) \\
{\cal M}(A^+,\phi^{-}_a,\phi^{-}_b,\phi^{+}_b,\phi^{+}_a)&=
\frac{\ab{24}\ab{53}}{\ab{51}\ab{12}\ab{34}}+(1-\lambda^2)\frac{\ab{52}}{\ab{51}\ab{12}}.
\end{split}
\end{equation}
Crucially, all of them are holomorphic in spinors. Under $[\lrbraces{1,2,3},0\rangle$ shift, it is straightforward to obtain the result by an MHV expansion from the above amplitudes~\cite{Cachazo:2004kj,Risager:2005vk}
\begin{equation}
	\begin{split}
		{\cal M}(\phi^-_a,\phi^-_b,\phi^-_c,\phi^+_c,\phi^+_b,\phi^+_a)
		=&
		\,\frac{[6\eta][\eta 1]}{[61]\langle 5 \slashed{P}_{61}\eta]\langle 2 \slashed{P}_{61}\eta]}
		\left(\frac{\ab{24}\ab{53}}{\ab{34}}+(1-\lambda^2) \ab{52} \right)\\
		& + \frac{[3\eta][\eta 4]}{[34]\langle 2 \slashed{P}_{34}\eta]\langle 5 \slashed{P}_{34}\eta]}
		\left(\frac{\ab{51}\ab{26}}{\ab{61}}+(1-\lambda^2) \ab{25} \right)\\
		& + \frac{1}{P_{612}^2}\left(\frac{\langle 1 \slashed{P}_{612}\eta]\ab{62}}{\langle 2 \slashed{P}_{612}\eta]\ab{61}}+(1-\lambda^2)\right)
		\left( \frac{\langle 4 \slashed{P}_{612}\eta]\ab{35}}{\langle 5 \slashed{P}_{612}\eta]\ab{34}}+(1-\lambda^2)\right) \\
		& + \frac{1}{P_{561}^2}\left(\frac{\langle 3 \slashed{P}_{561}\eta]\ab{24}}{\langle 2 \slashed{P}_{561}\eta]\ab{34}}+(1-\lambda^2)\right)
		\left( \frac{\langle 6 \slashed{P}_{561}\eta]\ab{51}}{\langle 5 \slashed{P}_{561}\eta]\ab{61}}+(1-\lambda^2)\right),
	\end{split}
\end{equation}
where $\eta$ is the reference spinor and $P_{\cal F}$ denotes the total momentum of the states in the factorization channel ${\cal F}$.
We have verified numerically that the answer is, as expected, independent of reference $\eta$.
Since the scalar amplitude is independent of the fermions, this result applies to any theory with the same bosonic sector.
When $\lambda=1$, the $SU(3)$ flavor symmetry together with the $U(1)$ $R$-symmetry combine to form the $SU(4)$ $R$-symmetry of ${\cal N}=4$ SYM.   Our expression agrees with known answer in this limit.

\begin{figure}[t]
	\centering
	\begin{subfigure}{0.3\textwidth}
		\hspace{0.5cm}
		\begin{fmffile}{general_scalar}
			\begin{fmfgraph}(100,100)
				\fmfleft{i1,i2,i3}
				\fmfright{o1,o2,o3}
				\fmf{dashes}{i1,v1}
				\fmf{dashes}{i2,v1}
				\fmf{dashes}{i3,v1}
				\fmf{dashes}{o1,v2}
				\fmf{dashes}{o2,v2}
				\fmf{dashes}{o3,v2}
				\fmf{dashes}{v1,v2}
			\end{fmfgraph}
		\end{fmffile}
		\vspace{0.5cm}
		\caption{general scalar.}
	\end{subfigure}
	\quad
	\begin{subfigure}{0.3\textwidth}
		\hspace{0.5cm}	
		\begin{fmffile}{charged_scalar}
			\begin{fmfgraph*}(100,100)
				\fmfleft{i1,i2,i3}
				\fmfright{o1,o2,o3}
				\fmf{dashes}{i1,v1}
				\fmf{dashes}{i2,v1}
				\fmf{dashes}{i3,v1}
				\fmf{dashes}{o1,v2}
				\fmf{dashes}{o2,v2}
				\fmf{dashes}{o3,v2}
				\fmf{dashes}{v1,v2}
				\fmflabel{$+$}{i1}
				\fmflabel{$-$}{i2}
				\fmflabel{$+$}{i3}
				\fmflabel{$-$}{o1}
				\fmflabel{$+$}{o2}
				\fmflabel{$-$}{o3}
			\end{fmfgraph*}
		\end{fmffile}
		\vspace{0.5cm}
		\caption{$U(1)$ charged scalar.}
	\end{subfigure}
	\quad
	\caption{Factorization channels in the 6-point scalar amplitude in $\phi^4$ theory.  The left and right diagrams show the factorization channels for the general case and the case of a $U(1)$ charged scalar, respectively. }
	\label{fig:scalar_channels}
\end{figure}
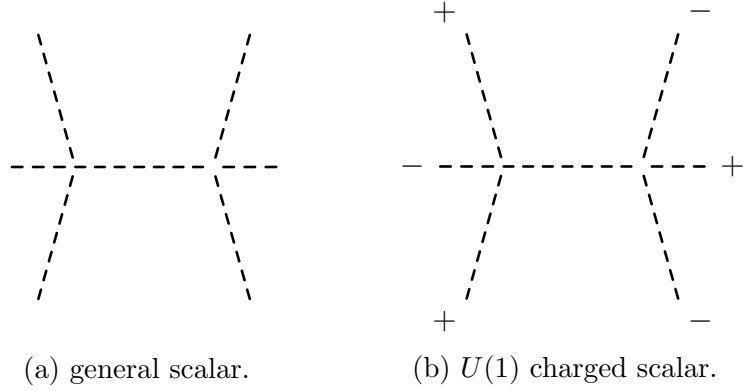

\paragraph{$\phi^4$ Theory.}
Next, consider amplitudes in a theory of interacting scalars.  We have shown that a 5-line shift is sufficient to construct all amplitudes, while a 3-line shift suffices if every scalar has equal charge under a $U(1)$ symmetry.  It is  straightforward to see how these apply to the 6-point scalar amplitude in $\phi^4$ theory. Applying a 5-line shift, the factorization channel is depicted in~\Fig{fig:scalar_channels}~where we  sum over all non-trivial permutations of external particles. If the scalar is complex and carries $U(1)$ charge, namely $|\phi|^4$ theory, then only channels satisfying charge conservation can appear.  Thus, three plus charged scalars never appear on one side of factorization. Consequently, shifting three plus charge scalars will construct the amplitude by exposing all physical poles.

\paragraph{$\psi^4$ Theory.}
From our previous discussion, we know four fermion theory can be constructed by a $Q^2\neq 0$ $5$-line shift. Consider a 6pt ${\cal M}(\psi^+,\psi^-,\psi^+,\psi^-,\psi^+,\psi^-)$ amplitude. Using a $[\lrbraces{2,4},\lrbraces{1,3,5}\rangle$ shift, we find
\begin{eqnarray}
	& & {\cal M}(\psi^+,\psi^-,\psi^+,\psi^-,\psi^+,\psi^-) \\
	&=& \sum_{{\cal P}(1,3,5),{\cal P}(2,4,6)} (-1)^\sigma \frac{[13]\ab{46}}{4P^2_{456}} \left ( \frac{z_{+,456}\langle 2| \hat{P}_{456} |5]|_{z_{-,456}}-(z_{+,456}\leftrightarrow z_{-,456})}{z_{+,456}-z_{-,456}} \right ) \\
	& = & \sum_{{\cal P}(1,3,5),{\cal P}(4,5,6)} (-1)^\sigma \frac{[13]\ab{46}\langle 2| P_{456} |5]}{4P^2_{456}},
\end{eqnarray}
where hatted variable is evaluated at factorization limit and $z_{\pm,456}$ are the two solutions of $\hat{P}^2_{456}=0$. The result is summed over permutation of $(1,3,5)$ and $(2,4,6)$ with $\sigma$ being the number of total permutation. In the last line, we use the fact that $\langle 2| \hat{P}_{456} |5]$ is linear in $z$ and only non-deformed part survives after exchanging $z_{\pm,456}$. We see the final answer has no square root as claimed before.

\paragraph{Maxwell-Einstein Theory.}
We discuss the theory where a $U(1)$ photon minimally couples to gravity. The coupling constant has the same dimension as in GR (see Tabel~\ref{table:coupling_summary}). But as a photon has less spin than a graviton, the large $z$ behavior is worse. We focus on the amplitudes with only external photons given that any amplitude with a graviton can be recursed by BCFW shift\cite{Cheung:2008dn}. Using a $m$-line $Q^2\neq0$ shift, we find $\Ms \lesssim z^{n+2-2m}$ at large $z$; thus, it's always possible to construct such an amplitude when $m>(n+2)/2$. Together with BCFW shift on gravitons, the theory is fully constructible! Using \Eqs{eq:bound_h}{eq:bound_mixed}, the result for $Q^2=0$ $m$-line shifts are
\begin{equation}
\gamma \leq \left\{ 
\begin{array}{ll}
1+n-3m/2,\quad & \text{for}\quad [\lrbraces{-,-,...},0\rangle\\
n-3m/2,\quad & \text{for}\quad [\lrbraces{-,-,...},\lrbraces{+}\rangle
\end{array}.
\right.
\end{equation}

For the 4pt ${\cal M}(A^-,A^-,A^+,A^+)$ amplitude,
we choose a $[\lrbraces{1,2};4\rangle$ shift so $\gamma<0$. The inputs for recursions are $3$pt functions obtained from consistency relation~\cite{Benincasa:2007xk}, ${\cal M}(A^-,A^+,G^-)=\ab{31}^4/\ab{12}^2$ and ${\cal M}(A^-,A^+,G^+)=[23]^4/[12]^2$. The amplitude then follows
\begin{eqnarray}
{\cal M}(A^-,A^-,A^+,A^+)=
\frac{\langle 1\hat{P}_{24}|4]^4}{\ab{13}^3[13][24]^2}\Big|_{z_{24}}
+\frac{\langle 2\hat{P}_{14}|4]^4}{\ab{23}^3[23][14]^2}\Big|_{z_{14}}
=\ab{12}^2[34]^2\left(\frac{1}{P^2_{24}}+\frac{1}{P^2_{14}}\right).
\end{eqnarray}

\paragraph{$F^v$ Operators.} Consider amplitudes with a single insertion of a $F^v$ operator. Applying a $[m,0\rangle$-line shift on minus helicity gluons and $[m-1,1\rangle$ $m$-line shift on all-but-one minus helicity gluons, \Eq{eq:bound_h} and \Eq{eq:bound_mixed} predicts
\begin{equation}
	\gamma \leq \left\{ 
		\begin{array}{ll}
		v-m,\quad & \text{for}\quad [\lrbraces{-,-,...},0\rangle\\
		v-1-m,\quad & \text{for}\quad [\lrbraces{-,-,...},\lrbraces{+}\rangle
		\end{array}.
	\right.
\end{equation}
We conclude $[v+1,0\rangle$- and $[v-1,1\rangle$-line shifts suffice to construct the amplitude with the given helicity configuration.

The case of $F^3$ operator has been studied extensively in  \Ref{Broedel:2012rc}. Given the large $z$ behavior above, the general MHV-like expression in \Ref{Dixon:2004za} can be proven inductively by a $[\lrbraces{-,-},\lrbraces{+}\rangle$ shift. In addition, the vanishing of boundary term in $[\lrbraces{-,-,...},0\rangle$ shift directly proves the validity of CSW-expansion in \Ref{Broedel:2012rc}. We demonstrate it with the MHV-like amplitude ${\cal M}(A^-,A^-,A^-,A^+)$ where a single $F^3$ operator is inserted. Note that the all-minus amplitude ${\cal M}(1^-,2^-,3^-)=\ab{12}\ab{23}\ab{31}$ is induced by a $F^3$ operator. Taking this as an input for the $[\lrbraces{2,3},4\rangle$ shift, we find
\begin{eqnarray}
	{\cal M}(A^-,A^-,A^-,A^+) &=& \ab{12}\ab{23}\ab{31}\left(\frac{\ab{23}}{\ab{34}\ab{2\hat{4}}}\big|_{z_{12}} -\frac{\ab{12}}{\ab{41}\ab{2\hat{4}}}\big|_{z_{23}}\right)\\
	&=& \frac{\ab{12}^2\ab{23}^2\ab{31}^2}{\ab{12}\ab{23}\ab{34}\ab{41}}.
	\label{eq:F3_amp}
\end{eqnarray}
This agrees with the result in \Ref{Dixon:1993xd,Dixon:2004za}.

The case of $\phi\,{\rm tr}(FF)$ operator, which is popular for the study of Higgs phenomenology, is very similar to $F^3$ operator. The MHV-like formula and CSW expansion in \Ref{Dixon:2004za} can also be proved analogously.

\paragraph{$R^v$ Operators.} Such operators often arise in effective theories from string action. Consider amplitudes with a single insertion of a $R^v$ operator. The amplitude scales as $z^{2v+n-3m}$ under a $m$-line $Q^2 \neq0$ shift. For a given $R^r$ operator, any $(n>v)$-pt amplitude can be constructed under an all-line $Q^2 \neq 0$ shift.  If we use $Q^2=0$ shifts, \Eq{eq:bound_h} and \Eq{eq:bound_mixed} give
\begin{equation}
\gamma\leq \left\{ 
\begin{array}{ll}
n+v-2m,\quad & \text{for}\quad [\lrbraces{-,-,...},0\rangle\\
n+v-2-2m,\quad & \text{for}\quad [\lrbraces{-,-,...},\lrbraces{+}\rangle
\end{array}.
\right.
\end{equation}
So if the helicity configuration is available, the amplitude is constructible under the $[m,0\rangle$- and $[m-1,1\rangle$-line shifts for $m>(n+v)/2$ and $m>(n+v)/2-1$, respectively.

Consider the 4pt ${\cal M}(G^-,G^-,G^-,G^+)$ amplitude with one $R^3$ operator insertion. We adopt the $[\lrbraces{2,3};4 \rangle$ shift to construct it. The amplitude factorizes into the anti-MHV amplitude in GR and ${\cal M}(G^-,G^-,G^-)=\ab{12}\ab{23}\ab{31}$ induced by one insertion of $R^3$ operator. We find
\begin{eqnarray}
	{\cal M}(G^-,G^-,G^-,G^+) &=& (\ab{12}\ab{23}\ab{31})^2\times\left[\frac{\ab{12}^2[41]}{\ab{\hat{4}2}^2\ab{41}}\Big|_{z_{41}}+(\text{cyclic in }(1,2,3)) \right], \\
	&=& (\ab{12}\ab{23}\ab{31})^2\left[
	\frac{[41]\ab{\xi 1}^2}{\ab{41}\ab{\xi 4}^2}+
	\frac{[42]\ab{\xi 2}^2}{\ab{42}\ab{\xi 4}^2}+
	\frac{[43]\ab{\xi 3}^2}{\ab{43}\ab{\xi 4}^2}
	\right] \\
	&=& P^2_{12}\,{\cal M}(1^-_A,2^-_A,4^+_A,3^-_A){\cal M}(1^-_A,2^-_A,3^-_A,4^+_A),
\end{eqnarray}
where $|\xi\rangle$ is a reference spinor in 3-line shift. The result in second line is manifest the leading soft factor of particle $4$. After canceling the reference spinor, the result in the last line is expressed in a KLT-relation form, where ${\cal M}(1^-_A,2^-_A,3^-_A,4^+_A)$ is the corresponding amplitude in gauge theory with $F^3$ operator given in \Eq{eq:F3_amp}. It agrees with \Ref{Broedel:2012rc}. It obvious from the answer that any $[m,0\rangle$ shift cannot construct the amplitude.

\section{Outlook}
\label{sec:outlook}

In this paper we have determined the minimal set of recursion relations needed to construct renormalizable and non-renormalizable field theories of massless particles in four dimensions.  We have shown that all renormalizable theories are constructible from a shift of five external momenta.  Quite surprisingly, a shift of three external momenta suffices for a more restricted but still enormous class of theories: all renormalizable theories in which the scalars, if present, are charged equally under a $U(1)$ symmetry.  Hence, we can construct all scattering amplitudes in any gauge theory with fermion and complex scalar matter, any supersymmetric theory, and the standard model.

Our results suggest several avenues for future work.   Because our analysis hinges solely on dimensional analysis, Lorentz invariance, and locality, it should be possible to generalize our approach to a broader class of theories.  In particular, there is the question of theories residing outside of four dimensions and involving massive particles.  Moreover, one might study an expanded covering space of recursion relations that include multiple complex deformation parameters or simultaneous shifts of holomorphic and anti-holomorphic spinors of the same leg.   

The recursion relations presented here might also offer new tools for studying the underlying properties of  amplitudes.  For example, the enhanced large $z$ behavior of amplitudes at large momenta implies so-called ``bonus relations'' whose nature remains unclear.  In addition, the soft shift defined in \Eq{eq:softshift} gives a nicely on-shell regulator for the soft limit of the amplitude.   Precise knowledge of the soft limit can uniquely fix effective theories \cite{Cheung:2014dqa}, and might actually be useful in the recursive construction of amplitudes, as we will discuss in \cite{Cheung:20xx}.  Finally, given a more complete understanding of on-shell constructibility at tree-level, we are better equipped to attack a much more difficult problem, which is developing a recursive construction for the loop integrands of general quantum field theories. This was accomplished for amplitudes in planar $\mathcal{N}=4$ SYM~\cite{ArkaniHamed:2010kv}, but with a procedure not obviously generalizable for less symmetric theories, where standard BCFW recursion induces ill-defined contributions in the forward limit. In principle, this somewhat technical obstruction might be eliminated by considering alternative momentum shifts.

\section*{Acknowledgments}
This research is funded by Walter Burke Institute for Theoretical Physics.
C.C. and C.-H.S. are supported by a DOE Early Career
Award under Grant No.~DE-SC0010255. C.C. is also supported by a Sloan Research Fellowship.
J.T. is supported in part by the David and Ellen Lee Postdoctoral Scholarship
and by the Department of Energy under grant number
DE-SC0011632.

\end{document}